\documentclass{aa}
\usepackage{epsfig}
\usepackage{txfonts}

\usepackage{epsfig}
\newcommand{\msun}{\mbox{$M_{\odot}$}}
\newcommand{\Msun}{\mbox{$M_{\odot}$}}

\newcommand{\teff}{\mbox{$T_{\rm eff}$}}

\newcommand{\kms}{km s$^{-1}$}

\begin{document}


\title{The nature of B supergiants: clues from a steep drop in rotation rates at 22\,000 K}

\subtitle{The possibility of Bi-stability braking}

\author{Jorick S. Vink\inst{1}, I. Brott\inst{2}, G. Gr\"afener\inst{1}, N. Langer\inst{3}, A. de Koter\inst{2,4}, D.J. Lennon\inst{5} }
\offprints{Jorick S. Vink, jsv@arm.ac.uk}

 \institute{Armagh Observatory, College Hill, Armagh, BT61 9DG, Northern Ireland
            \and
Astronomical Institute, Utrecht University, Princetonplein 5, 3584 CC, Utrecht, The Netherlands
\and
Argelander-Institut f\"ur Astronomie der Universit\"at Bonn, Auf dem H\"ugel 71, 53121 Bonn, Germany
\and
Astronomical Institute Anton Pannekoek, University of Amsterdam, Kruislaan 403, 1098 SJ, Amsterdam, The Netherlands
\and
ESA, Space Telescope Science Institute, 3700 San Martin Drive, Baltimore, MD 21218, USA
}

\titlerunning{Rotation and bi-stability}
\authorrunning{Jorick S. Vink}

\abstract{The location of B supergiants in the Hertzsprung-Russell 
diagram (HRD) represents a long-standing problem in massive star evolution. 
Here we propose their nature may be revealed utilising their 
rotational properties, and we highlight a steep drop in massive star 
rotation rates at an effective temperature of 22\,000 K.
We discuss two potential explanations for it. On the one hand, the
feature might be due to the end of the main sequence, which could 
potentially constrain the core overshooting parameter. On the other hand, the feature might be 
the result of enhanced mass loss at the predicted 
location of the bi-stability jump. We term this effect ``bi-stability braking'' and 
discuss its potential consequences for the evolution of massive stars. 
}

\maketitle


\section{Introduction}
\label{s_intro}

The large number of B supergiants as well as their location in the Hertzsprung-Russell 
diagram represents a long-standing problem in massive star evolution (e.g. Fitzpatrick \& Garmany 
1990). Even the most basic question of whether B supergiants are core hydrogen (H) 
burning main sequence (MS) or helium burning objects has yet to be answered. 
Here we propose their nature may be revealed utilising their rotational properties. 

On the MS, O-type stars are the most rapid rotators known 
(with $\varv$$\sin$$i$ up to 400 \kms), but B supergiants rotate much more slowly 
(with $\varv$$\sin$$i$ $\la$ 50\kms), which has been attributed to the expansion of the 
star after leaving the MS.
Hunter et al. (2008) noted a steep drop in rotation rates at low gravities 
(log $g$ $<$ 3.2) and suggested the slowly rotating B supergiants to be post-MS. 
The steep drop was also used to constrain the core overshooting parameter $\alpha_{\rm ov}$ 
in massive star models (Brott et al. 2010). 
The slowly rotating B supergiants are also cooler (with $\teff$ below $\sim$22\,000 K) 
and $\varv$$\sin$$i$ is observed to drop steeply below this $\teff$.  
Here we introduce an alternative explanation for the slow rotation of B 
supergiants: wind-induced braking due to bi-stability, or bi-stability braking (BSB).

Mass loss plays a crucial role in the evolution of massive stars. 
Whilst a large amount of attention has been directed towards the role of stellar winds in terms 
of the {\it loss of mass}, as winds ``peel off'' the star's outer layers (Conti 1976), 
much less effort has been dedicated to understanding the associated {\it loss of angular momentum} 
(but see Langer 1998, Meynet \& Maeder 2003). Yet  
the angular momentum aspect of these winds may be equally relevant for understanding 
massive stars as the loss of mass itself, possibly in a mass range as low as $\sim$10-15\msun.

We first recapture the physics of bi-stable winds and BSB (Sect.~\ref{s_bsb}), before 
presenting the current knowledge of rotational velocities of massive stars. We note 
a steep drop at $\sim$22\,000 K (Sect.~\ref{s_hook}) and propose two possible explanations for it.
In the first one, the drop is due to the separation of MS objects from a second population
of slow rotators (Sect.~\ref{s_eoms}), whilst in the second one the slow rotation 
is the result of BSB (Sect.~\ref{s_cbsb}).


\section{The physics of the mass loss bi-stability jump} 
\label{sec_intro2}

The BS-Jump (Pauldrach \& Puls 1990) is a theoretically predicted 
discontinuity where wind properties change from a modest $\dot{M}$, fast wind, 
to a higher $\dot{M}$, slow wind, when the effective temperature drops below $\sim$22\,000 K.
Vink et al. (1999) predicted an increase in the mass-loss rate by a factor of $\sim$5-7 
(see the blue dotted line in Fig.~\ref{f_brott})
and a drop in the terminal wind velocity by a factor of $\sim$2. Here the reason for the jump 
is an increased flux-weighted effective number of iron lines due to the recombination of Fe {\sc iv} to 
Fe {\sc iii} (and not necessarily related to the optical depth of the Lyman continuum). 
In fact, the temperature of the BS-Jump was found to be weakly density dependent, with 
the BS-Jump starting as high as $\sim$26 kK for the higher mass models 
and dropping to $\sim$22.5 kK for the lower mass models at $\sim$20\Msun\ (Vink et al. 2000). 

Whilst the predicted drop in terminal wind velocity across the BS range 
has been confirmed (Crowther et al. 2006), the issue of a jump in mass loss 
is controversial. The jump may have been confirmed in 
radio data that suggest a local $\dot{M}$ maximum at the 
predicted temperature (Benaglia et al. 2007), but the predicted rates on the cool side of 
the jump are up to an order of magnitude larger than those found from state-of-the-art NLTE 
models (Vink et al. 2000, Trundle \& Lennon 2005, Crowther et al. 2006, Markova \& Puls 2008, 
Searle et al. 2008).
To gain more insight into this mass-loss discrepancy one way forward
is to search for other physical effects in the bi-stability region, which 
might assist us to unravel whether B supergiant mass-loss rates are as high 
as predicted, or as low as the spectral modelling suggests. 
Here we outline one such approach, involving stellar rotation rates in the region
of the BS-Jump. Vink (2008) pointed out that the temperature of the bi-stability jump coincides 
with the position where the rotational velocities 
drop to smaller values (below 100 km/s), and suggested this might be due to BSB.

\subsection{Bi-stability braking (BSB)}
\label{s_bsb}

\begin{figure}
\centerline{\psfig{file=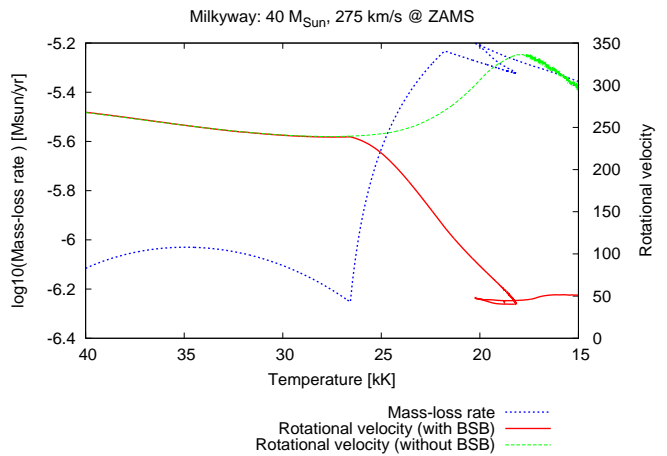, angle=270,width = 8.7 cm}}
\caption{Mass-loss rate (blue dotted) and rotational velocities for a 40 \msun\ star
with an initial rotational velocity of 275 \kms, including the predicted 
BS-Jump (red solid) and without it (dashed green).
A core overshooting parameter $\alpha_{\rm ov}$ of 0.335 was employed in these models.}
\label{f_brott}
\end{figure}

Employing stellar evolution models which include mass loss and rotation, we test whether the predicted 
increase in $\dot{M}$ at the bi-stability jump might lead to slower rotation.
Figure~\ref{f_brott} shows the run of the mass-loss rate and predicted rotational velocity as a function of
temperature for a 40 \msun\ star with an initial rotational velocity of 275 \kms. 
It shows a drastic drop in surface rotation rates for massive stars around 22\,000 K, which is 
due to BSB in our models. When we do not increase the mass loss due to the BS-Jump, 
the stars remain rotating rapidly -- despite the stellar expansion, as   
angular momentum is transferred from the core to the envelope.
Bi-stability braking can only be efficient if the star spends a significant amount of time, i.e. part of its MS evolution, on 
the cool side of the BS-Jump. In our present standard models BSB only occurs above 
a critical mass of $\sim$30 \msun\ for the Galaxy, $\sim$35 \msun\ for the Large Magellanic Cloud (LMC), 
and $\sim$50 \msun\ for the Small Magellanic Cloud (SMC). In these models we employed a core 
overshooting parameter $\alpha_{\rm ov}$ of 0.335 of a pressure scale-height. 
A higher value of $\alpha_{\rm ov}$ lowers the critical mass. For instance, in a 
test calculation with $\alpha_{\rm ov}$=0.5 at LMC metallicity, BSB occurs already for a 20$\msun$ star.

\section{The steep drop in rotational velocity at 22\,000 K}
\label{s_hook}

\begin{figure}
\centerline{\psfig{file=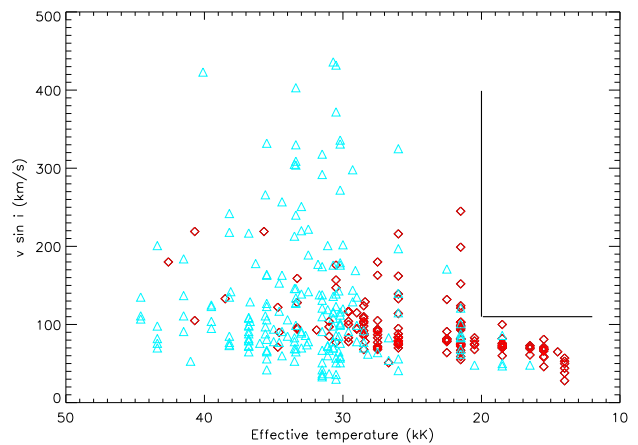, width = 8.7 cm}}
\caption{Projected rotational velocity $\varv$$\sin$$i$ of the Howarth et al. (1997) 
dataset of Galactic OB supergiants (red diamonds) and non-supergiants (blue triangles) 
as a function of $\teff$ (converted from spectral types using 
Martins et al. 2005 and Crowther et al. 2006). 
We note that $\varv$$\sin$$i$ drops from values as high as 
$\sim$400/250\kms\ to values below 100 \kms\ at $\sim$22\,kK.}
\label{f_howarth}
\end{figure}

Howarth et al. (1997) catalogued $\varv$$\sin$$i$ values for 373 
OB stars, with roughly half of them being supergiants (luminosity class I; red diamonds). 
We plot the $\varv$$\sin$$i$ values of this large and uniformly determined data-set in 
Fig.~\ref{f_howarth}. The figure shows a drop in $\varv$$\sin$$i$ for stars 
hotter than 22 kK with values as high as $\sim$400 km/s for all objects and as high 
as $\sim$250 km/s for the supergiants only to values that {\it all} fall 
below 100 km/s for the cooler objects. In other words, we identify a general 
absence of rapidly rotating B supergiants\footnote{Note that the remaining broadness in the B supergiant spectra may (partly) be due 
to macro-turbulence in addition to, or instead of, rotational broadening (Conti \& Ebbets 1977, Aerts et al. 2009).}. 
As the stars in this data-set have not been analysed in detail, we resort to 
the results from the {\sc flames} Survey of massive stars (Evans et al. 2008), which involves data 
from the Galaxy, the LMC, and the SMC. 

\begin{figure}
\centerline{\psfig{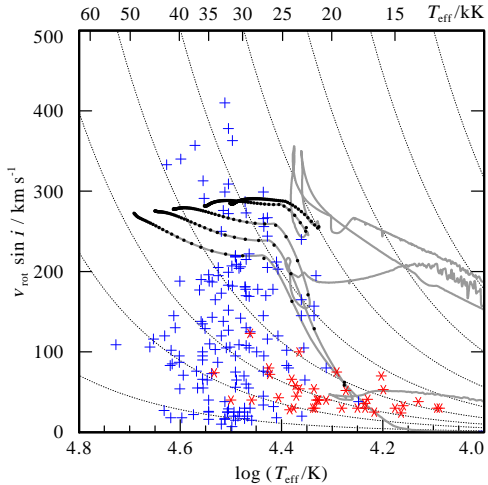}}
\caption{Rotational velocities vs. $\teff$ for all {\sc flames} objects with evolutionary 
masses above 15\,$M_\odot$. Luminosity classes are shown as blue pluses (luminosity classes {\sc ii-v}) 
and red stars (luminosity class {\sc i}). The LMC evolutionary tracks including the predicted BS-Jump are 
shown in grey with initial $\varv_{\rm rot}=250$\,km/s for five masses of 15, 20, 30, 40 and 60\,$M_\odot$. 
It can be noted that the critical mass for BSB is $\sim$35\msun\ in the LMC.
The steepness of these tracks can be compared to the angular momentum conservation case, drawn 
as grey dotted background lines.
The black dots on the tracks represent $10^5$ year time-steps.}
\label{VROT_GG}
\end{figure}

Figure~\ref{VROT_GG} shows $\varv$$\sin$$i$ versus effective temperature for the {\sc flames} data, where 
we again note a steep drop in $\varv$sin$i$ from values as high as $\sim$400\kms\ to values below 100\kms. 
The data selection is a non-trivial undertaking, as we wish to optimise the sample size to sample 
homogeneity -- in the presence of selection effects. 
The reason we employ a cut-off mass at 15\msun\ is that 
for the largest subset, i.e. that of the LMC, there is a detection limit that runs from $\sim$20\msun\ 
at the hottest temperatures to $\sim$10\msun\ at the cool part of the HRD (see Brott et al. 2010). 
For this reason, we choose an intermediate value of 15 \msun\ as a minimum value. 
If we had chosen a higher mass cut-off, the drop feature would shift to a somewhat higher $\teff$ (up to 27 kK). 
If we had opted for a lower mass cut-off, the feature shifts to $\teff$ $=$ 20 kK. Having noted this, tens of manual 
trials have shown that the sheer drop feature itself does not depend on a particular 
choice of mass range, and given the presence of the drop in both Figs.~\ref{f_howarth} and~\ref{VROT_GG}, 
as well as data presented in Fraser et al. (2010), we argue the drop feature is ubiquitous in the 
entire mass range $\sim$10-60$\msun$.

The dark grey lines overplotted in Fig.~\ref{VROT_GG} show evolutionary tracks for the intermediate 
metallicity of the LMC with $\varv_{\rm rot}=250$\,km/s by Brott et al.\,(2010) for masses of 
15, 20, 30, 40 and 60\,$M_\odot$.
The black tick-marked dots on the tracks 
represent evolutionary time-steps of $10^5$ years, and are intended 
to facilitate the comparison with observations.

\begin{figure}
\centerline{\psfig{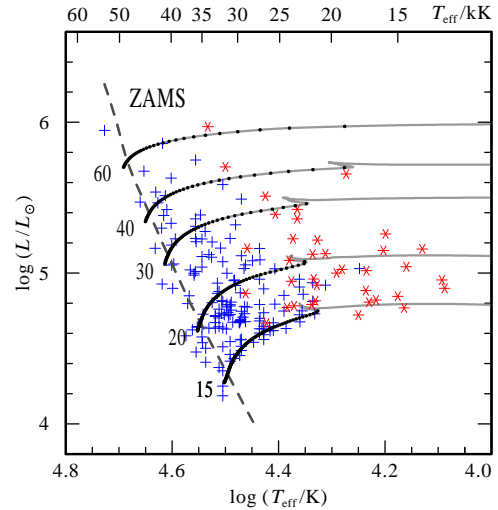}}
\caption{HRD of the {\sc flames} survey of massive stars. See the caption of Fig.\,\ref{VROT_GG} for an 
explanation of the symbols.}
\label{HRD_GG}
\end{figure}

\begin{figure}
\centerline{\psfig{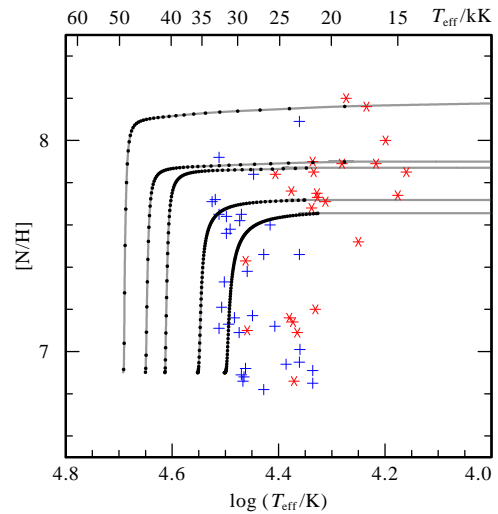}}
\caption{Nitrogen abundance vs. $\teff$ for the LMC subset. 
See the caption of Fig.\,\ref{VROT_GG} for an explanation of the symbols.}
\label{nitrogen}
\end{figure}

\section{Two possible interpretations for the drop in $\varv$sin$i$}
\label{s_inter}

In Sect.~\ref{s_hook}, we highlighted the steep drop in  
the rotation rates of massive stars, but we have yet to provide an explanation for it. 
The question is whether the absence of rapidly rotating B supergiants is the result of BSB, or if 
the cooler slow rotators form an entirely separate population from the hotter MS stars.

\subsection{The case for two populations}
\label{s_eoms}

The cool objects (red asterisks) in Fig.\,\ref{VROT_GG} and the HRD of Fig.~\ref{HRD_GG} are supergiants 
of luminosity class {\sc i}, whilst the fast rotators are
dominated by dwarfs (blue pluses). 
Although it is by no means obvious that supergiants 
cannot be in a H burning phase, the division 
in log$g$ might imply that we have a 
population of rapidly rotating MS objects on the one hand, whilst observing a 
population of slowly rotating evolved supergiants -- which have somehow lost their angular 
momentum -- on the other hand.
Currently, we do not have sufficient information with respect to 
the evolutionary state of these cool supergiants. In principle, this part of the HRD 
can be populated with the products of binary evolution, although this would normally not 
be expected to lead to slowly rotating stars. Alternatively, one 
could envision the cooler objects to be the product of single star evolution, e.g.  
post-RSG or blue-loop stars, but the key 
point is that within the context of the two population interpretation, they are 
{\it not} core H burning.

A potential distinguishing factor between the ``two population scenario'' and BSB 
is that of the chemical abundances.
We present LMC N abundances versus effective temperature in Fig.~\ref{nitrogen}, noting 
that N abundances could only be derived for a subset
of our objects shown in Figs.~\ref{VROT_GG} and~\ref{HRD_GG}. 
As the LMC baseline [N/H] equals $\sim$6.9, the vast majority of slow rotators is found 
to be strongly N enhanced. 
Although rotating models can in principle account for large N abundances, the fact that 
such a large number of the cooler objects is found to be N enriched suggests an 
evolved nature for these stars.

\subsection{The case for BSB}
\label{s_cbsb}

The second explanation for the steep drop in rotation rates 
is that both the objects cooler and hotter than 22\,000 K reside on the MS, and 
that it is BSB that explains the slow rotation of the cooler B supergiants.
The main argument for BSB is that it is predicted at the temperature
where the rotational velocities are found to drop steeply. 
The evolutionary tracks in Fig.\,\ref{HRD_GG} indicate that the
MS for the highest mass stars indeed appears rather broad, 
reaching as far as the BS-Jump temperature at 22\,kK, and beyond.   
Therefore, mass loss seems capable of removing a 
considerable amount of angular momentum during the MS evolution for the highest mass stars. 

\section{Discussion}
\label{s_disc}

In principle it is possible that both effects of ``two populations'' 
and BSB occur simultaneously at 22\,000 K, with BSB occurring above 
a certain critical mass, and the ``two population'' scenario
taking over in the lower mass (10-20 \msun) range, but this situation might
appear somewhat contrived. 
The strongest argument for the ``two population scenario'' are the large N abundances 
of the B supergiants, whilst the strongest argument for BSB is that the drop is 
observed at the correct location (whilst no such coincidence would be expected for 
the alternative interpretation).

Using our standard models, BSB can only operate above a certain
critical mass and would not be able to explain the steep drop in rotational velocities 
of stars below the critical mass. 
The reason BSB does not operate at lower masses in our standard models (of Fig.\,\ref{VROT_GG}) 
is that the drop feature has been used to constrain the core 
overshooting parameter of $\alpha_{\rm ov}$ = 0.335. 
The applicability of BSB could be pushed to lower masses if the 
MS lifetime were extended. This could 
be achieved by increasing $\alpha_{\rm ov}$. When we enlarge $\alpha_{\rm ov}$ to 0.5, 
BSB also occurs at 20\msun\ for our Galactic and LMC models. 
What is clear is that the critical mass is model-dependent. For instance,  
the solar-metallicity models of Meynet \& Maeder (2003) show BSB in the 
lower ($\sim$15-20\msun) range.

We point out that if BSB were the correct explanation
for the drop feature all the way down to $\sim$10$\msun$, we would
require a very large core overshooting parameter, and the consequences would 
be far-reaching. For instance, it would imply that B (and even A) supergiants are MS objects burning H in their cores.
This would potentially solve the long-standing problem of the presence of such a 
large number of B supergiants. Moreover, if BSB could work for the entire mass range, it would 
also have profound implications for the Blue to Red (B/R) supergiant ratio that has been 
used to constrain massive star models as a function of metallicity for decades.
Furthermore, if the absence of rapidly rotating B supergiants is due to BSB, one might
wonder what this would imply for the evolutionary state of the presumably 
rapidly rotating B[e] supergiants. The rapid rotation of these extreme 
objects could possibly be related to close binary evolution or merging 
(Pasquali et al. 2000), but this requires future investigation.

If BSB would indeed occur in the lower mass range (down to $\sim$10$\msun$), one 
should be aware that the derived overshooting parameter of 0.335 becomes a 
lower limit and that the real value becomes larger. 
Although this would be consistent with the suggested increase 
in $\alpha_{\rm ov}$ with stellar mass (Ribas et al. 2000), such a large
value of $\alpha_{\rm ov}$ might be considered uncomfortable, as the highest mass 
data-point in Ribas et al. is based on one binary star, V380 Cyg, for which the 
results have been challenged (Claret 2003). 

To summarise, we have presented two potential explanations 
for the steep drop in rotation rates at 22\,000 K. Currently, we 
have insufficient information to decide which one is correct. 
In any case, our study demonstrates the important role of mass loss for
massive star evolution, and especially the importance of {\it specifics} 
in its dependence on the stellar parameters. 
Furthermore, we have highlighted the significant influence of mass loss
on the angular momentum transport in massive stars.   
Last but not least, BSB may offer a novel method of diagnosing 
the effects of mass loss via its influence on the angular momentum.
Current analyses yield controversial results with respect to the existence of a 
BS jump. On the one hand, the predicted drop in terminal wind velocity across the BS range 
has been confirmed (Crowther et al. 2006). On the other hand, for temperatures 
below the BS-Jump, the mass-loss rates obtained from spectral modelling are generally 
much lower than predicted (Vink et al. 2000, Crowther et al. 2006).

A simultaneous investigation of the abundances, mass loss, and rotational 
properties of a large sample of massive stars, e.g. with the {\sc flames ii} 
Tarantula survey (Evans et al. 2009), would be most helpful to settle these issues.


\end{document}